\documentclass[aps,twocolumn,prd,showpacs,amsmath,amssymb]{revtex4}
\usepackage{dcolumn}
\usepackage{bm}

\newcommand{\be}{\begin{equation}}
\newcommand{\ee}{\end{equation}}
\newcommand{\ba}{\begin{eqnarray}}
\newcommand{\ea}{\end{eqnarray}}
\newcommand{\ban}{\begin{eqnarray*}}
\newcommand{\ean}{\end{eqnarray*}}
\newcommand{\n}[1]{\label{#1}}

\newcommand{\eq}[1]{(\ref{#1})}
\newcommand{\R}{\ensuremath{{\cal R}}}
\newcommand{\rhom}{\ensuremath{{\rho^m}}}
\newcommand{\rhot}{\ensuremath{{\rho^T}}}

\newcommand{\D}{\ensuremath{\tilde{\nabla}}}
\newcommand{\dm}{\ensuremath{\Delta_m}}
\def\la {\langle}
\def\ra {\rangle}

\begin{document}

\title{The Existence of  Einstein Static Universes and their Stability \\ in Fourth order Theories of Gravity}
\author{Rituparno Goswami$^\dag$, Naureen Goheer$^\dag$  and Peter K S 
Dunsby$^{\dag\diamond}$}
\affiliation{$^\dag$ Department of Mathematics and Applied\
Mathematics, University of Cape Town, South Africa.}
\affiliation{$^\diamond$ South African Astronomical Observatory,
Observatory Cape Town, South Africa.}

\begin{abstract}
We investigate whether or not an Einstein Static universe is a solution
to the cosmological equations in $f(R)$ gravity.  It is found that
only one class of $f(R)$ theories admits an Einstein Static model, and that
this class is neutrally stable with respect to vector and tensor perturbations
for all equations of state on all scales. Scalar perturbations are
only stable on all scales if the matter fluid equation of state
satisfies $c_{{\rm s}}^{2}>\frac{\sqrt{5}-1}{6}\approx 0.21$. This result is
remarkably similar to the GR  case, where it was found that
the Einstein Static model is stable for $c_s^2>\frac{1}{5}$.
\end{abstract}
\pacs{04.50.+h, 04.25.Nx }
\maketitle
\section{Introduction}
Although modifications of Einstein's theory of  gravity were already proposed in the early years after the publication of General Relativity (GR), a detailed investigation of cosmological models within this framework only got underway a few years ago. Such models became popular in the 1980's because it was shown that they naturally admit a phase of accelerated expansion which could be associated with an early universe inflationary phase \cite{star80}. The fact that the phenomenology of Dark Energy requires the presence of a similar phase (although only a late time one) has recently revived interest in these models. In particular, the idea that Dark Energy may have a geometrical origin, i.e., a that there is a connection between Dark Energy  and a non-standard behavior of gravitation on cosmological scales is currently a very hot topic of research.

In the last few years cosmologies based on fourth order gravity have been widely studied using several different, but complementary techniques.  A number of extremely interesting results have been found \cite{revnostra,Odintsov,Carroll} which suggest that it might be possible that these models provide a viable alternative to Dark Energy \cite{Capozziello:2005ku,Capozziello:2006dj,Nojiri:2007uq,Hu:2007nk}. Among these approaches,  a particularly interesting one is based on the analysis of the phase space of these models \cite{cdct:dynsys05,SanteGenDynSys,shosho,naureen1,naureen2}, providing a  systematic method for obtaining exact solutions together with their stability and a general idea of the qualitative behavior of these cosmological models.

Of particular interest is the fact that there are classes of $f(R)$  theories which admit a Friedmann  transient  matter-dominated decelerated expansion phase, followed by one with an accelerated expansion rate \cite{cdct:dynsys05,Capozziello:2006dj}. The first phase provides a setting during which structure formation can take place and this is followed by a smooth transition to a Dark Energy like era which drives the cosmological acceleration. It would therefore be of great interest if orbits could be found in the phase-space of $f(R)$ models that connect such a Friedmann matter dominated phase to an accelerating phase via an Einstein Static solution in a way which is indistinguishable (at least at the level of the background dynamics) to what occurs in the $\Lambda$CDM cosmology of GR \cite{Goliath}.  If a similar evolution exists in an $f(R)$ model, we would expect the associated Einstein Static solution to be a saddle point (as it is in GR) and consequently unstable.

The issue of stability of the Einstein Static model may at first seem a somewhat trivial matter. Indeed, already in the 1930s, Eddington \cite{eddington} showed that in GR such models are unstable with respect to homogeneous and isotropic perturbations, and ever since then the Einstein Static model has been considered unstable to gravitational collapse or expansion. Indeed,  it is exactly this feature that allows the transition between a decelerated expansion era to one which is accelerating in the $\Lambda$CDM cosmology. However, later work first by Harrison \cite{harrison} and then by Gibbons \cite{gibbons} suggests that this issue requires further investigation, both in the context of GR and extended theories of gravity. In Harrison's classic paper, he investigated the normal-modes of vibration of the universe and showed that all physical inhomogeneous modes are oscillatory in a radiation-filled Einstein Static model, while Gibbons showed that an Einstein Static model filled with a perfect fluid was stable with respect to conformal metric perturbations, provided that the sound speed satisfies $c_{{\rm s}}^{2} >{\frac{1}{5}}$.  The reason for this ``non-Newtonian" stability stems from the fact that in GR,  the Einstein Static universe is spatially closed therefore has a maximum scale associated with it, which is greater than the largest physical scale of the perturbations. Since the Jean's length is a significant fraction of this maximum scale, perturbations in the fluid oscillate, rather than grow, leading to conclusion that in GR at least, the Einstein Static solution is marginally stable with respect to such perturbations. This result was also found more recently by Barrow {\it et. al.} \cite{barrow03}, when linear inhomogeneous and anisotropic perturbations of the Einstein Static model were investigated.

Another reason why Einstein Static solutions and their stability are both interesting and important is due to the fact that a number of authors have conjectured that  the universe might have started out in an asymptotically Einstein Static state, providing a natural beginning to the inflationary expansion phase of the Universe \cite{emergant}.

The issue of stability also plays an important role in the viability of cosmological solutions  in $f(R)$ theories \cite{Faraoni:2006sy,Nojiri:2006ri,Sokolowski:2007pk,Amendola:2007nt}. Recently \cite{Sawicki:2007tf} it was argued that the sign of the second derivative
$f_{,RR}\equiv\partial^2f/\partial R^2$ determines whether the theory approaches the general relativistic limit at high curvatures, and it was shown
that for $f_{,RR}> 0$ the models are, in fact, stable. The stability of the de Sitter solution in $f(R)$ gravity has also been extensively analyzed in the literature \cite{deSitter}.

Motivated by the above discussion, the goal of this paper is to first determine under what conditions an Einstein Static solution exists in a general $f(R)$ gravity theory and investigate the stability of such solutions with respect to general inhomogeneous and isotropic perturbations.
\section{Einstein-Static Universes in Fourth order Gravity}
In a completely general context, a fourth order theory of gravity is obtained by adding terms involving
$R_{ab}R^{ab}$ and $R_{abcd}R^{abcd}$ to the standard Einstein Hilbert action.  However, it is now
well known that  if we use the Gauss Bonnet theorem we can neglect the $R_{abcd}R^{abcd}$
term \cite{GB}. Furthermore, if we  take into account the high symmetry of the FRW metric, the 
Lagrangian can be further simplified. Specifically  the variation of the term  $R_{ab}R^{ab}$ can always 
be rewritten in terms of the variation of $R^2$ \cite{Rsquared}. It follows that the "effective" fourth-order Lagrangian in FRW cosmology only contains powers of $R$ and we can, with out loss of generality, write the action as 
\be
{\cal A}=\int d^4x\sqrt{-g}\left[f(R)-2\Lambda+{\cal L}_m\right]\;,
\n{action}
\ee
where ${\cal L}_m$ represents the matter contribution and $\Lambda$ is the usual cosmological constant.
The corresponding generalization of Einstein's equations are
\be
G_{ab}+g_{ab}\frac{\Lambda}{f'}=T_{ab}^T=\frac{T_{ab}^m}{f'}+
T_{ab}^R\;,
\n{EE1}
\ee
where $f'=f(R)_{,R}$. Here  $T_{ab}^T$ is the total effective energy momentum tensor, $T_{ab}^m$ is the
energy momentum tensor for standard matter and
\be
T_{ab}^R=\frac{1}{f'}\left[\frac{1}{2}g_{ab}(f-Rf')+
f'_{;cd}(g^c_ag^d_b-g^{cd}g_{ab})\right]
\n{TR}
\ee
is the energy momentum tensor containing all the non-GR curvature contributions, known as
the "{\it curvature fluid}".

The metric for a closed Einstein Static universe is given by
\be
ds^2=-dt^2+a_0^2\left[\frac{dr^2}{1-r^2}+r^2d\Omega^2\right]\;.
\n{metric}
\ee
Here $a_0$ is a constant and both the Ricci scalar $R$ and the
curvature of the 3-space $\tilde{R}$ are equal and can be
written as
\be
R=\tilde{R}=\frac{6}{a_0^2}\;.
\n{Ricci}
\ee
It is easily seen that an Einstein Static universe is expansion, shear
and rotation free, i.e.
\be
\Theta=0,\;\sigma_{ab}=0,\;\omega_{ab}=0\;.
\ee
The total energy momentum tensor is given by
\ba
\rhot=\frac{1}{f'}\left[\rhom+\frac{1}{2}(Rf'-f)\right]\nonumber\\
p^T=\frac{1}{f'}\left[p^m-\frac{1}{2}(Rf'-f)\right]\n{T}\;,
\ea
where $\rho_m$ and $p_m$ is the energy density and pressure of standard matter.

Now the $\left(^t_t\right)$ component of Einstein's equations gives
\be
\frac{R}{2}=\rhot+\frac{\Lambda}{f'}\n{tt}\;,
\ee
while the  trace of Einstein's equations is
\be
R-4\frac{\Lambda}{f'}=\rhot-3p^T\;.
\n{trace}
\ee
Furthermore, from the Raychaudhuri equation we have $R_{ab}u^au^b=0$, from which we
obtain
\be
\rhot+3p^T=2\frac{\Lambda}{f'}\;.
\n{ray}
\ee
Using \eq{tt} and \eq{ray} we find
\be
R=3\left[\rhot+p^T\right]\;.
\n{R1}
\ee
Now using the components of total energy momentum tensor
\eq{T}, and assuming that the standard matter is a barotropic perfect
fluid with equation of state and sound-speed parameters $w$ and $c^2_s$ defined by
\be
p^m=w\rhom,\;\;-1\le w\le1\;,~~ c_s^2=w\;,
\n{eos}
\ee
equation \eq{R1} becomes
\be
Rf'=3(1+w)\rhom\;.
\n{R2}
\ee
The LHS of the above equation cannot be zero for this spacetime,
because this would imply either $a_0\rightarrow\infty$ or the function $f(R)=constant$.
Consequently, we cannot have an Einstein Static universe without matter or with $w=-1$.

It follows from \eq{R2} above, together with the trace equation \eq{trace}
and \eq{T} that for an Einstein Static universe to exist, the function $f(R)$ needs
to satisfy the following differential equation:
\be
Rf'-\frac{3}{2}(1+w)f+3\Lambda(1+w)=0\;.
\n{diff1}
\ee
In order to obtain the GR limit we set $f=R$ and $f'=1$. In this case we obtain
\be
R=\frac{6}{a_0^2}=\frac{6\Lambda(1+w)}{(1+3w)}\;.
\n{gr}
\ee
It is easy to see that in GR, once we fix $a_0$ and $\Lambda$,
both $w$ and $\rhom$ get fixed via equations \eq{gr} and
\eq{R2}.
The general solution for \eq{diff1} is given by
\be
f(R)=2\Lambda+{\cal K}R^{\frac{3}{2}(1+w)}\;,
\n{gensol}
\ee
where ${\cal K}$ is the constant of integration.
It follows that the action for the Einstein Static universe \eq{action}
becomes
\be
{\cal A}=\int d^4x\sqrt{-g}\left[{\cal K}R^{\frac{3}{2}(1+w)}
+{\cal L}_m\right]\;.
\n{action1}
\ee
We can then easily see that in fourth order gravity,  the existence of an Einstein Static universe implies there is
no explicit dependence on the cosmological constant in the action. However, we know
that in GR an Einstein Static universe exists for the action
\be
{\cal A}=\int d^4x\sqrt{-g}\left[R-2\Lambda+{\cal L}_m\right]\;,
\n{action2}
\ee
with the Ricci scalar given by \eq{gr}. 
As noted earlier, the basic difference between GR and fourth order theories of gravity, 
in case of Einstein Static Universe is, for GR fixing $w$ and $\Lambda$ would fix 
$\rhom$ and $a_0$ simultaneously. Whereas in $f(R)$ gravity fixing the former quantities 
would fix the `function' $f$ and not the values of $\rhom$ and $R$. Hence we can in principle 
have an Einstein Static Universe with any non-zero positive $\rhom$ for a given barotropic 
index and cosmological constant and this is a direct 
consequence of the extra degrees of freedom for higher order gravity. However if we wish 
to impose the condition that for the value of the Ricci scalar given by \eq{gr}, the 
function $f(R)$ also has the same value (so that we have a GR limit), that would then  
fix the constant of integration
${\cal K}$ as
\be
{\cal K}=\frac{4\Lambda}{(1+3w)}\left[\frac{(1+3w)}
{6\Lambda(1+w)}\right]^{\frac{3}{2}(1+w)}\;.
\n{K}
\ee
However, if we switch off the cosmological constant term in
\eq{action} from the very beginning,  it is possible to get an Einstein Static
universe for any value of ${\cal K}$.

It is extremely interesting to note that just the background symmetry of the Einstein Static universe fixes the form
of allowed actions in the fourth order gravity theory.  It follows that,  based on the prescription leading to equation (14), many commonly used forms of  fourth order gravity, such as  $f(R)=R+\alpha R^n$ or  $f(R)=R+\alpha/R$ do not admit an have an Einstein Static solution.

\section{Stability analysis}

In the previous section we described the complete dynamical equations  describing an Einstein Static
universe in fourth order gravity and discovered that there is a unique class of $f(R)$ theories that admits
such a solution. In what follows, we investigate the stability of this model with respect to generic linear
inhomogeneous and anisotropic perturbations by linearizing the most general propagation and constraint
equations for this $f(R)$ theory around the Einstein Static background. This is done using the
1+3 covariant approach to perturbations \cite{fRpert1,fRpert2,BED,BDE,DBE,DBBE}, where quantities that vanish
in the background spacetime are considered to be first order and are automatically gauge-invariant by virtue of the
Stewart and Walker lemma \cite{SW}. In order to write down the set of linear equations we first
need to choose a physically motivated frame $u^a$. The natural choice is the one which is tangent to the matter flow lines.

The projection tensor into the tangent 3-spaces orthogonal to this flow vector is
\be
h_{ab}=g_{ab}+u_au_b\;.
\ee
Also, the totally projected covariant derivative operator
$\D_a$, orthogonal to $u^a$ is defined by
\be
\D_a=h^b_a\nabla_b\;.
\ee
The effective total density and pressure in the presence of perturbations is then
\be
\rhot=\frac{1}{f'}\left[\rhom+\frac{1}{2}(Rf'-f)+f''\D^2R\right]\;,
\n{rhot1}
\ee
\be
p^T=\frac{1}{f'}\left[p^m-\frac{1}{2}(Rf'-f)-
\frac{2}{3}f''\D^2R+f''\ddot{R}\right]\;,
\n{pt1}
\ee
while the anisotropic part of effective energy momentum tensor
is given by
\be
\Pi^T_{ab}=\frac{1}{f'}\left[f''\D_{<a}\D_{b>}R\right]\;,
\n{pi}
\ee
and the energy flux is
\be
q^T_a=\,-\frac{1}{f'}\left[f'''\dot{R}\D_{a}R+f''\D_{a}\dot{R}-\frac{1}{3}f''
\D_{a}R\right]\;.
\ee
Here angle brackets denote the projected trace-free part (see \cite{cargese} for details).

In terms of these effective thermodynamical quantities,
the perturbed cosmological equations are
\be
\dot{a}=\frac{1}{3}\Theta a\;,
\n{adot}
\ee
\be
\Theta^2=3\left[\rhot+\frac{\Lambda}{f'}\right]
-\frac{3}{2}\tilde{R}\;.
\n{theta}
\ee
In addition to these, energy conservation for
standard matter is given by
\be
\dot{\rhom}=-\Theta\rhom(1+w)\;,
\n{conserv}
\ee
while the momentum conservation gives the following relation
connecting the acceleration to the matter density:
\be
A^a=\dot{u}^a=-\frac{w}{w+1}\frac{\D^a\rhom}{\rhom}\;.
\n{acc}
\ee
The expansion propagation (or the generalized Raychaudhuri equation) is
\be
\dot{\Theta}=-\frac{1}{3}\Theta^2+\D^aA_a
-\frac{1}{2}\left(\rhot+3p^T\right)+\frac{\Lambda}{f'}\;.
\n{ray1}
\ee
The shear propagation equation can be written as
\be
\dot{\sigma}_{ab}=-\frac{2}{3}\Theta\sigma_{ab}-E_{ab}
+\frac{1}{2}\Pi_{ab}+\D_{\la a}A_{b\ra}\;,
\n{sigmadot}
\ee
where the tensor $E_{ab}$ is the "electric" part of the
Weyl tensor. The gravito-electric propagation equation
is given by
\ba
\dot{E}_{ab}=&-\Theta E_{ab}+curl(H_{ab})
-\frac{1}{2}\left(\rhot+p^T\right)\sigma_{ab}\nonumber\\
&-\frac{1}{6}\Theta\Pi_{ab}-\frac{1}{2}\dot{\Pi}_{ab}-\frac{1}{2}\D_{\la a}q_{b\ra}\;,
\n{edot}
\ea
where $H_{ab}$ is the "magnetic" part of Weyl tensor.
The gravito-magnetic propagation equation is
\be
\dot{H}_{ab}=-\Theta H_{ab}-curl(E_{ab})+\frac{1}{2}curl(\Pi_{ab})\;.
\n{hdot}
\ee
The system of equations is closed with the vorticity propagation equation which
governs rotational perturbations:
\be
\dot{\omega}_a=-\frac{2}{3}\Theta\omega_a-\frac{1}{2}curl(A_a) \;.
\n{omega}
\ee
In what follows we will use the following linearized identities for
closed Einstein static background, for any scalar $g$, vector
$V^a$ and tensor $A_{ab}$:
\begin{eqnarray}
(\D^ag)\,\dot{}=\D^a(\dot{g})\,,\\
(\D^aV^b)\,\dot{}=\D^a(\dot{V}^b)\,,\\
(\D^2g)\,\dot{}=\D^2(\dot{g})\,,\\
\D_a(\D^2g)=\D^2(\D_ag)-\frac{2}{a_0^2}(\D_ag)
\end{eqnarray}
and
\be
curl(curl A_{ab})=-\D^2A_{ab}+\frac{3}{a_0^2}A_{ab}
+\frac{3}{2}\D_{<a}\D^cA_{b>c}\,.
\ee
Finally, to convert the system of partial differential equations
to ordinary differential equations, we use the standard
procedure of harmonic decomposition, employing the trace-free
symmetric tensor eigenfunctions of the spatial Laplace-Beltrami
operator defined by
\be
\D^2Q=-\frac{k^2}{a_0^2}Q\;,\; \dot{Q}=0.
\ee
We can then expand all first order quantities (both scalar and tensor)
as
\be
X(t,{\bm{{\rm x}}})=\sum X^k(t)Q^k({\bm{{\rm x}}})
\ee
For the spatially closed models, the spectrum of eigenvalues is
discrete and given by \cite{harrison,BDE}
\be
k^2=n(n+2)\;,
\ee
where the co-moving wavenumber $n=1,2,3,\cdots$.
The mode $n=0$ corresponds to a change in the background. The
first inhomogeneous mode $n=1$ is a gauge mode which is
ruled out by the constraint equations $G_{0i}=T_{0i}$, and reflects the freedom
to change the 4-velocity of the fundamental observers. It follows that  physical
modes are constrained by $k^2\ge8$  \cite{harrison,BDE}.

In the following sections we will extract the scalar, vector
and tensor components from the linearized perturbation equations
using the standard procedure \cite{BDE} and analyze the stability of
the Einstein Static universe.

\subsection{Scalar Perturbations}
In order to derive the equations governing scalar perturbations, we define
variables describing the spatial Laplacians of density,  expansion and spatial curvature
\be
\dm=\frac{a^2\D^2\rhom}{\rhom},\;Z=a^2\D^2\Theta,\;
C=a^4\D^2\tilde{R}\;,
\n{var1}
\ee
and the dimensionless Laplacians describing the
inhomogeneity in the Ricci scalar and its momentum
\be
\R=a^2\D^2R\;,\; \Re=a^2\D^2\dot{R}\;.
\n{var2}
\ee
In what follows we obtain the evolution equations for
the above scalar variables. Using the linearized identities
together with \eq{conserv} we get
\be
\dot{\Delta}_m=-(1+w)Z\;.
\n{deldot}
\ee
The generalized Raychaudhuri equation \eq{ray1} and
the trace equation \eq{trace} can be combined to eliminate the $\ddot{R}$
term, giving
\ba
\dot{Z}=&-&\left[\frac{R}{3(1+w)}\right]\dm+\frac{1}{2}\R\nonumber\\
&-&\frac{w}{w+1}\D^2\dm-\left[\frac{1+3w}{2R}\right]\D^2\R\;.
\n{zdot}
\ea
Again from the linearized identities we get
\be
\dot{\R}=\Re\;,
\n{rdot}
\ee
and using \eq{trace} we have
\ba
\dot{\Re}&=&\frac{2R^2}{9}
\left[\frac{1-3w}{(1+w)(1+3w)}\right]\dm\nonumber\\
&+&\frac{R}{3}\left[\frac{3w-1}{3w+1}\right]\R + \D^2\R
\n{rdotdot}
\ea
Finally the constraint and the propagation equations for
the curvature perturbation $C$ are
\be
\frac{C}{a^2}=\left[\frac{2R}{3(1+w)}\right]\dm
+\left[\frac{1+3w}{R}\right]\D^2\R
\n{c}
\ee
and
\be
\frac{\dot{C}}{a^2}=-\left[\frac{2R}{3}\right]Z
+\left[\frac{1+3w}{R}\right]\D^2\Re\;.
\n{cdot}
\ee
The system of equations \eq{deldot}-\eq{cdot} are coupled
partial differential equations. We use harmonic
decomposition to convert them into a system of coupled
first order ordinary differential equations. Following \cite{fRpert1} we can then
write them as a pair of coupled second order equations:
\ba
\ddot{\Delta}^k_m=\left[\frac{2-k^2w}{a_0^2}\right]
\dm^k-\left[\frac{(1+w)(6+k^2+3wk^2)}{12}\right]\R^k\;,\nonumber\\
\n{delddot}
\ea
\ba
\ddot{\R}^k&=&\left[\frac{8(1-3w)}{a_0^4(1+4w+3w^2)}\right]
\dm^k\nonumber\\
&+&\left[\frac{6w-2wk^2-2-k^2}{a_0^2(1+3w)}\right]\R^k\;,
\n{rddot}
\ea
where we have used $R=6/a_0^2$. The above equations can be
combined to a single fourth order equation for $\dm^k$:
\be
\frac{d^4}{dt^4}\dm^k -A^k(w)\frac{d^2}{dt^2}\dm^k
+B^k(w)\dm^k=0\;,
\n{delfour}
\ee
where
\be
A^k(w)=-\frac{3k^2w^2+4k^2w+k^2-12w}{a_0^2(1+3w)}\;
\n{A}
\ee
and
\be
B^k(w)=\frac{k^2(9k^2w^2+3k^2w-36w^2-12w-4)}{3a_0^4(1+3w)}.
\n{B}
\ee
The general solution for equation \eq{delfour} can be easily obtained as
\be
\dm^k(t)=C_1e^{-\beta_-t}+C_2e^{+\beta_-t}
+C_3e^{-\beta_+t}+C_4e^{+\beta_+t}\;,
\n{solution}
\ee
where
\be
\beta_{\pm}=\sqrt{2A^k(w)\pm2\sqrt{A^k(w)^2-4B^k(w)}}\;.
\n{beta}
\ee
It is easy to see that the solution \eq{solution} is non-growing if and only
if all the conditions
\begin{enumerate}
\item $A^k(w)<0$
\item $B^k(w)>0$
\item $A^k(w)^2-4B^k(w)>0$
\end{enumerate}
are satisfied, and it follows that all three conditions
are satisfied for all $k\in[8,\infty)$ if
\be
w>\frac{-1+\sqrt{5}}{6}\equiv w_c\;.
\ee
In other words,  the Einstein Static universe is stable with respect to
all scalar perturbations for $w_c<w<1$. This range includes
radiation and stiff matter, but not pressure free matter. It is also interesting
that the ratio of $w_{rad}$ and $w_c$ is the {\it golden ratio}.
\section{Vector Perturbations}
The vector perturbations are sourced by the vorticity tensor
$\omega_{ab}$. At linear order, the vorticity propagation
equation for the Einstein Static background is simply
\be
\dot{\omega}_{ab}=0\;.
\ee
It follows that the Einstein Static model is neutrally stable
against vector perturbations for all equations of state, on all scales.
\section{Tensor Perturbations}
Tensor perturbations \cite{fRpert2} are characterized by the evolution of the transverse,
trace-free part of the shear tensor $\sigma_{ab}$. At linear
order the shear propagation equation decouples:
\be
\ddot{\sigma}_{ab}=\D^2\sigma_{ab}+
\frac{1}{2}(\rho^T+p^T-R)\,\sigma_{ab}\;,
\ee
where we have only kept the tensor contributions.
Substituting the exact solution for the Einstein Static background
and decomposing into Fourier modes with co-moving index $k$,
we obtain
\be
\ddot{\sigma}_{ab}+\frac{k^2+2}{a_0^2}\sigma_{ab}=0\;.
\ee
Solving this equation, it can easily be seen that for all equations of state, the
Einstein Static model is neutrally stable against tensor perturbations on all scales.
\section{Conclusions}
In this work, we have given a careful analysis of  the stability of  Einstein Static universe
in fourth order theories of gravity with respect to generic inhomogeneous and anisotropic perturbations.

First of all, we showed that the modified field equations only admit an Einstein Static solution for the very special form
of gravitational Lagrangian: $f(R)=a+bR^c$, where the constants $a,~b,~c$ are fixed functions of equations of state
parameter $w$ and cosmological constant $\Lambda$ if $\Lambda\neq0$. If $\Lambda=0$,  $a=0$ and $b$ is arbitrary.
This is a surprising result, since one could naively expect that the additional degrees of freedom arising from a modified $f(R)$
would make it easier to find solutions of the Einstein Static type.  This result  differs from some recent work \cite{portsmouth}, where the
stability of the Einstein Static solution was investigated for a number of types of $f(R)$ theories which appear not to fall into the
class discovered here.

Our stability analysis shows that as in GR \cite{barrow03}, the Einstein static model is neutrally stable against vector and tensor perturbations for all equations of state at all scales. Scalar perturbations are only stable on all scales if the matter fluid equation of state satisfies $c_s^2>\frac{\sqrt{5}-1}{6}\approx 0.21$. This result is remarkably similar to the GR case, where it was found that the Einstein Static model is stable for $c_s^2>\frac{1}{5}$.
\acknowledgments
We thank Sante Carloni and Kishore Ananda for valuable input and the National Research Foundation (South Africa) for financial support.

\end{document}